\documentclass[preprint,12pt]{elsarticle}
\usepackage[left=3cm, right=3cm, top=3cm, bottom=3cm]{geometry}
\usepackage[utf8]{inputenc}

\usepackage{textcomp}

\usepackage{amsmath}
\usepackage{amsfonts}
\usepackage{graphicx}
\usepackage{amssymb}
\usepackage{subcaption}

\usepackage{lineno}

\usepackage{booktabs} 

\usepackage{tikz}
\usetikzlibrary{arrows}

\usepackage{algorithm}
\usepackage[noend]{algpseudocode}
\algnewcommand\algorithmicforeach{\textbf{for each}}
\algdef{S}[FOR]{ForEach}[1]{\algorithmicforeach\ #1\ \algorithmicdo}
\algnewcommand{\algorithmicand}{\textbf{ and }}
\algnewcommand{\algorithmicor}{\textbf{ or }}
\algnewcommand{\OR}{\algorithmicor}
\algnewcommand{\AND}{\algorithmicand}
\algnewcommand{\var}{\texttt}

\newcommand{\semres}[5]{$\chi^2 \geq #1$, $\mathrm{df} = #2$, $p \geq #3$, $\mathrm{RMSEA} \leq #4$, $\mathrm{SRMR} \leq #5$}

\journal{Biomedical Signal Processing and Control}

\begin{document}


\begin{frontmatter}

\title{Automating LC--MS/MS mass chromatogram quantification: Wavelet transform based peak detection and automated estimation of peak boundaries and signal-to-noise ratio using signal processing methods.}

\author[TUD]{Florian Rupprecht}
\author[TUD,MSB]{Sören Enge}
\author[TUD]{Kornelius Schmidt}
\author[TUD]{Wei Gao}
\author[TUD]{Clemens Kirschbaum}
\author[PFI,TUD]{Robert Miller}

\address[TUD]{Technische Universität Dresden (TUD), Department of Psychology, Dresden, Germany}
\address[MSB]{Medical School Berlin (MSB), Department of Psychology, Faculty of Science, Berlin, Germany}
\address[PFI]{Health Technology Assessment \& Outcomes Research, Pfizer Germany GmbH, Linkstr. 10, 10785 Berlin, Germany}

\begin{abstract}

\textbf{Background and Objective:} 
While there are many different methods for peak detection, no automatic methods for marking peak boundaries to calculate area under the curve (AUC) and signal-to-noise ratio (SNR) estimation exist.
An algorithm for the automation of liquid chromatography tandem mass spectrometry (LC--MS/MS) mass chromatogram quantification was developed and validated. 
\textbf{Methods:}
Continuous wavelet transformation and other digital signal processing methods were used in a  multi-step procedure to calculate concentrations of 6 different analytes. 
To evaluate the performance of the algorithm, the results of the manual quantification of 446 hair samples with 6 different steroid hormones by two experts were compared to the algorithm results.
\textbf{Results:}
The proposed approach of automating LC--MS/MS mass chromatogram quantification is reliable and valid. The algorithm returns less non-detectables than human raters. 
Based on signal to noise ratio, human non-detectables could be correctly classified with a diagnostic performance of AUC = $0.95$.
\textbf{Conclusions:}
The algorithm presented here allows fast, automated, reliable, and valid computational peak detection and quantification in LC--MS/MS.

\end{abstract}


\end{frontmatter}



\section{Introduction}

A very time consuming part of LC--MS/MS quantification and related methods is manual peak marking in the raw mass chromatograms.
While many different methods for peak detection have been developed \citep{yang2009pdalgcomparison}, to our knowledge, there is no open source method which also is able to mark peak boundaries for area under the curve (AUC) measurement and perform signal-to-noise ratio (SNR) estimation.
This article describes an automated procedure capable of performing these tasks, handles fluctuations in retention time (RT) and perform calibration to a known reference standard.

There are three direct advantages of analyzing LC--MS/MS mass chromatograms by implementing an automated process:
Increased productivity and speed by decreasing manual labour, increased objectivity and reliability resulting out of the absence of human raters and, resulting from the non-proprietary nature of the described method, increased replicability~\citep{plesser2018reproducibility}.

This article first describes underlying concepts and models applied by the algorithm. Then a test with real world data is performed and a comparison of the algorithm results to results manually quantified by human experts is made.


Finally the following two hypotheses were evaluated:

\begin{enumerate}
    \item The algorithm returns analyte concentrations which correlate strongly with those manually quantified by experts.
    \item The algorithm returns analyte concentrations which are valid.
\end{enumerate}

The manuscript is structured as follows. In section~\ref{sec:alg} the algorithm and its theoretical background are described, in section~\ref{sec:exp} algorithm results are compared to results manually compiled by human raters.

\section{Theory and calculation}
\label{sec:alg}

The following sections describe an algorithm for automatic LC--MS/MS mass chromatogram quantification. The first section introduces a generic model of chromatogram composition which forms a theoretical basis for the subsequent sections about decomposition, transformation and processing of measured chromatograms. The final sections then outline a multi-step procedure that is necessary to calculate the absolute concentrations of an analyte.

\subsection{A generic model of mass chromatograms}

\begin{figure}[ht]
    \centering
    \includegraphics[width=0.5\textwidth]{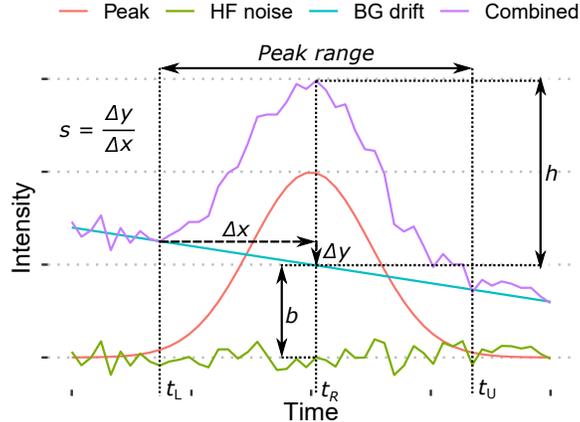}
    \caption{Generic model for chromatographic peaks. The actual LC--MS/MS sensor data ($y(t)$) is combined background and drift ($B(t)$), peak distribution ($P(t)$) and noise ($N(t)$). Note that this Figure does not include peak skew or nearby peaks. Peak range is the time interval between the lower bound ($t_\mathrm{L}$) and upper bound ($t_\mathrm{U}$). Retention time (RT, $t_\mathrm{R}$) indicates maximum sensor intensity in the peak range. Peak height ($h$) is background ($b = B(t_\mathrm{R})$) subtracted by sensor intensity of the combined signal at RT $y(t_\mathrm{R}) - b$. Slope ($s$) is the slope of the background component. Note that in the implementation background and slope are defined using the line between lower and upper bounds of the peak range to simplify the process. This is not necessarily equivalent to the definition in this model.}
    \label{fig:chromacomp}
\end{figure}

A chromatogram can be described as a time series of LC--MS/MS detector intensities $y(t) \in \mathbb{R}$, where $t \in \mathbb{R^+}$ indicates time. The relevant sequence containing the peak can be constructed from several component series using the following additive model.

\begin{equation}
y(t) = B(t) + P(t) + N(t) \,,
\end{equation}

where $B(t) = b + (t - t_R) \cdot d$ is a linear trend component containing background $b \in \mathbb{R^+}$ and linear drift $d \in \mathbb{R}$ of the chromatogram, $P(t)$ is the peak distribution component, and $N(t)$ is an irregular component containing noise (besides low frequency background noise, this also can contain nearby peaks and low frequency noise or drift). 
These components are illustrated in Figure~\ref{fig:chromacomp}. 

Retention time (RT, $t_R$) is the time at which the peak component has its maximum intensity. Peak height ($h$) is background ($b = B(t_\mathrm{R})$) subtracted by  signal intensity at RT $y(t_\mathrm{R}) - b$. Slope ($s$) is the slope of the background component. All these measures are annotated in Figure~\ref{fig:chromacomp}.

To measure analyte concentration, the area of the peak needs to be calculated. 
Peak components are determined by performing several data transformations on the chromatogram time series.

\subsection{Data transformations}
\label{sec:trans}

\begin{figure}[!ht]
    \centering
    \includegraphics[width=0.6\textwidth]{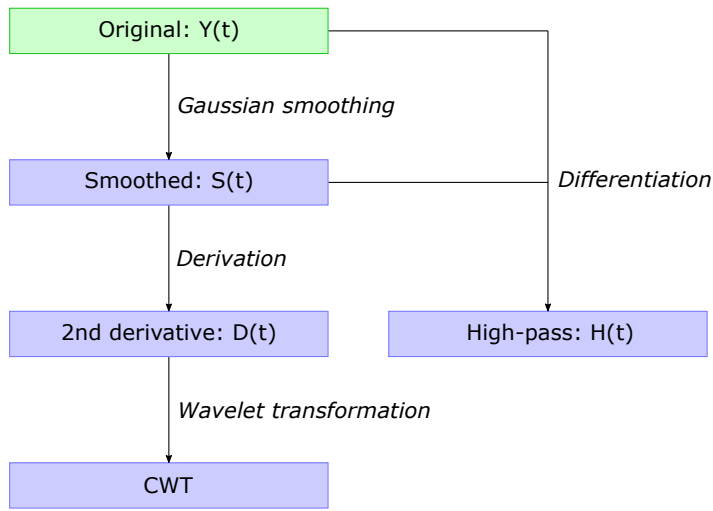}
    \caption{Chromatogram time series transformations. The LC--MS/MS chromatographic time series $Y(t)$ is smoothed using a gaussian kernel. High-pass filtered data is obtained by differentiating smoothed and original time series. The second derivative of the smoothed data is approximated and the continuous wavelet transformation (CWT) is calculated.}
    \label{fig:dataflow}
\end{figure}

The following number of transformations are calculated for a given chromatogram.
Figure~\ref{fig:dataflow} illustrates the transformations and their order.

\paragraph{Smoothing / low-pass filtering}
The data are smoothed using a 1-D gaussian kernel smoother \citep[see][p. 40-42]{davies2012computer}. The following formula returns aggregation weights by which chromatogram intensity values are smoothed. Smoothness can be adjusted using the manually set hyperparameter $\sigma$.

\begin{equation}
G(t) = \frac{1}{\sqrt[]{2 \pi } \sigma} \exp{\left(- \frac{t^{2}}{2 \sigma ^{2}}\right)}
\label{eq:1dgaussian}
\end{equation}

\paragraph{High-pass filtering}
By differentiation of the original $Y(t)$ and smoothed $S(t)$ time series, the high-pass filtered data $H(t)=Y(t)-S(t)$ is received. This makes it dependent on the $\sigma$ parameter of the gaussian kernel.

\paragraph{Derivation}
The second derivative $D(t) = S''(t)$ of the smoothed data $S(t)$ is approximated by applying 

\begin{equation}
    S'\left(\frac{t_i+t_{i-1}}{2}\right) = \frac{S(t_i)-S(t_{i-1})}{t_i-t_{i-1}}
    \label{eq:diff}
\end{equation} 

twice.

\paragraph{Wavelet decomposition}

Continuous wavelet transformations are used to create time-frequency breakdowns of the chromatograms \citep[see][]{daubechies1992wavelets}.

Exponentially modified Gaussian distributions are commonly used to model chromatographic peaks \citep{Grushka_1972}. The second derivative of the gaussian distribution function is used to build a mother wavelet. Chromatographic peaks have a lower frequency than the interfering noise. This enables the separation and further analysis of the noise and peak distributions by investigating CWT frequency breakdowns.

Continuous wavelet transforms are used to divide the data\footnote{The original mass chromatogram is unevenly spaced. While the transformations up to this point can handle this type of data, the continuous wavelet transform implementation is constrained to the use of evenly sampled data.
The data are then resampled in even time steps using linear interpolation. If data loss is a concern the sampling resolution can be increased.} into wavelets. 
The continuous wavelet transform of a function $x(t)$ with
scale $a \in \mathbb{R^+}$ and a translation variable $b\in \mathbb{R}$ can be expressed by the following integral:

\begin{equation}
X_w(a,b)=\frac{1}{\sqrt{a}}\int_{-\infty}^\infty x(t)\psi\left(\frac{t-b}{a}\right)dt
\end{equation}

Chromatographic peaks are commonly modelled by exponentially modified gaussian distributions \citep{Grushka_1972}. The negative normalized second derivative of a Gaussian function is used. It is usually called the \emph{mexican hat wavelet} \citep{torrence1998practical}:

\begin{equation}
\psi(t) = \frac{2}{\sqrt{3\sigma}\pi^{1/4}} \left(1 - \left(\frac{t}{\sigma}\right)^2 \right) \exp{\left(-\frac{t^2}{2\sigma^2}\right)}
\end{equation}

Note that the processed sensor data also has to be differentiated so that the Gaussian derivative wavelet fits to the peak distribution.
This furthermore has the advantage, that both background and linear drift are eliminated from the raw mass chromatogram.

\begin{figure}[ht]
    \centering
    \includegraphics[width=0.7\textwidth]{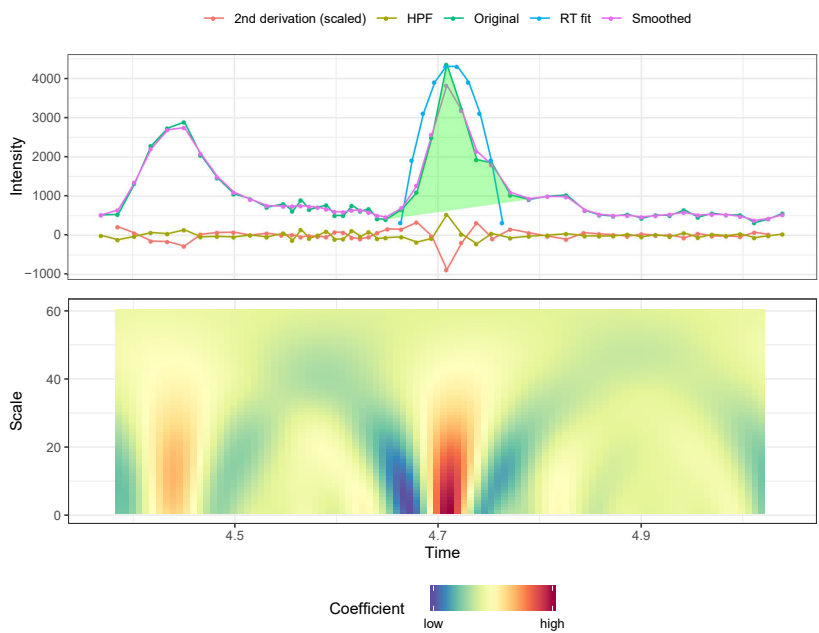}
    \caption{Chromatographic time series, second derivative (scaled for visibility), and the resulting coefficient matrix after wavelet transformation. The time-intensity graph also contains smoothed and high-pass filtered (HPF) time series, as well as relevant points of the retention time fitness function (RT fit) which is referred to in section~\ref{sec:peakselection}.}
    \label{fig:cwtexample}
\end{figure}

Then the continuous wavelet transform is used to construct a time-frequency representation of a signal. Figure~\ref{fig:cwtexample} shows a chromatographic time series, its second derivative, and the resulting CWT coefficient matrix.

\subsection{Chromatogram processing}
\label{sec:chromaproc}

Each single chromatogram is analyzed in multiple steps. 
Possible peaks, their RTs, and area boundaries are identified first. Then signal-to-noise ratio (SNR), as well as peak area, height, background and slope are calculated. Lastly a fitness value is calculated by which the peak most likely associated with the analyte is selected.

\subsubsection{Peak identification \& boundary estimation}
\label{sec:alg:ppos}

\begin{figure}[!ht]
    \centering
    \begin{subfigure}[t]{0.3\textwidth}
        \includegraphics[width=\textwidth]{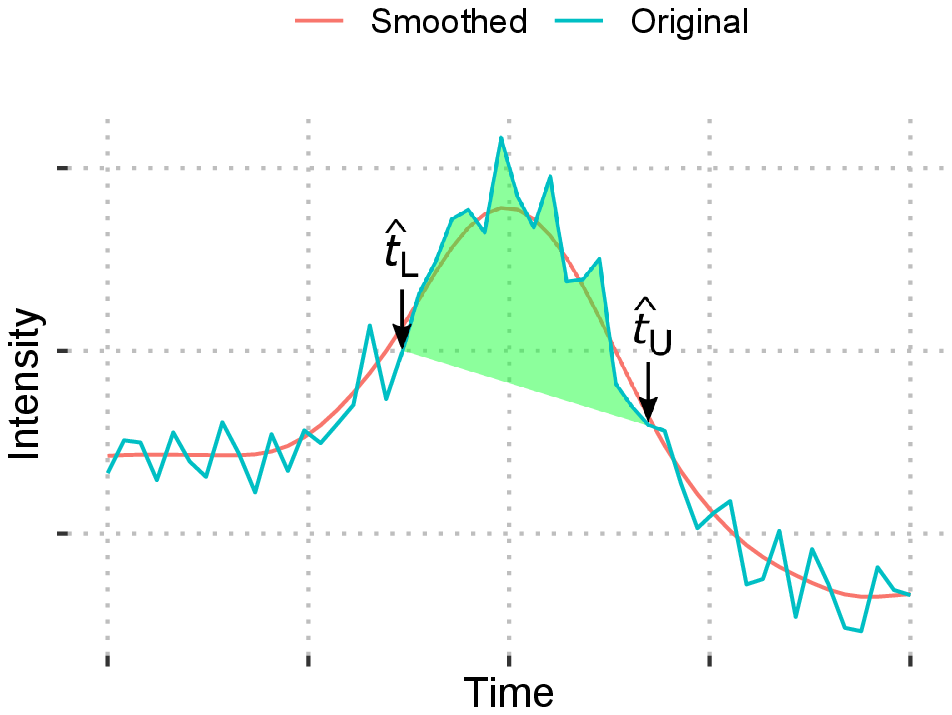}
        \caption{Peak area after initial boundary estimation.}
        \label{fig:peakboundssteps1}
    \end{subfigure}
    ~~
    \begin{subfigure}[t]{0.3\textwidth}
        \includegraphics[width=\textwidth]{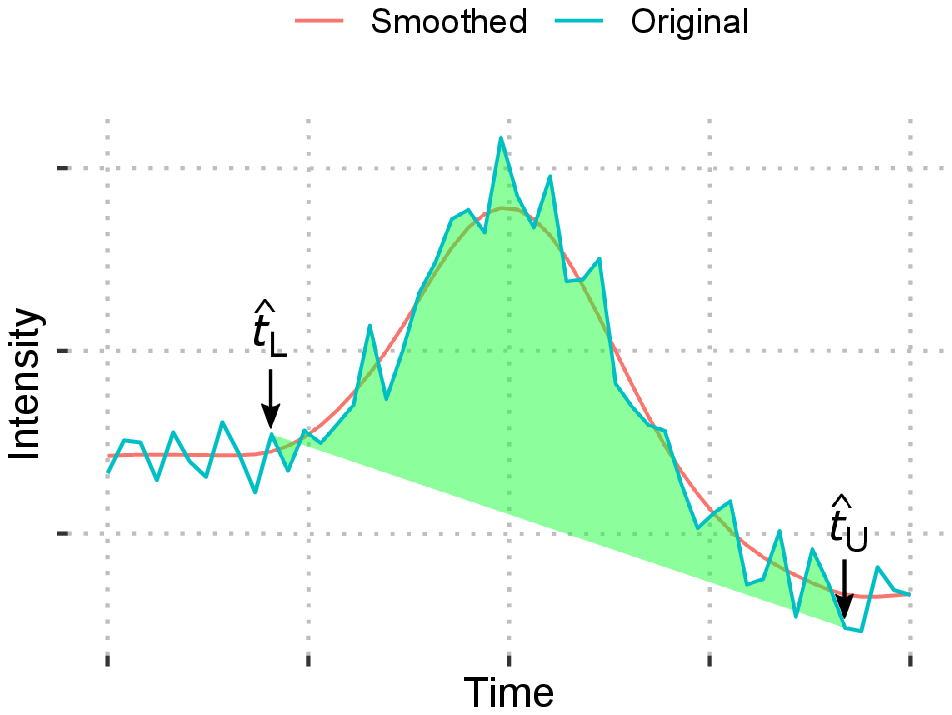}
        \caption{Peak area after friction boundary correction.}
        \label{fig:peakboundssteps2}
    \end{subfigure}
    ~~
    \begin{subfigure}[t]{0.3\textwidth}
        \includegraphics[width=\textwidth]{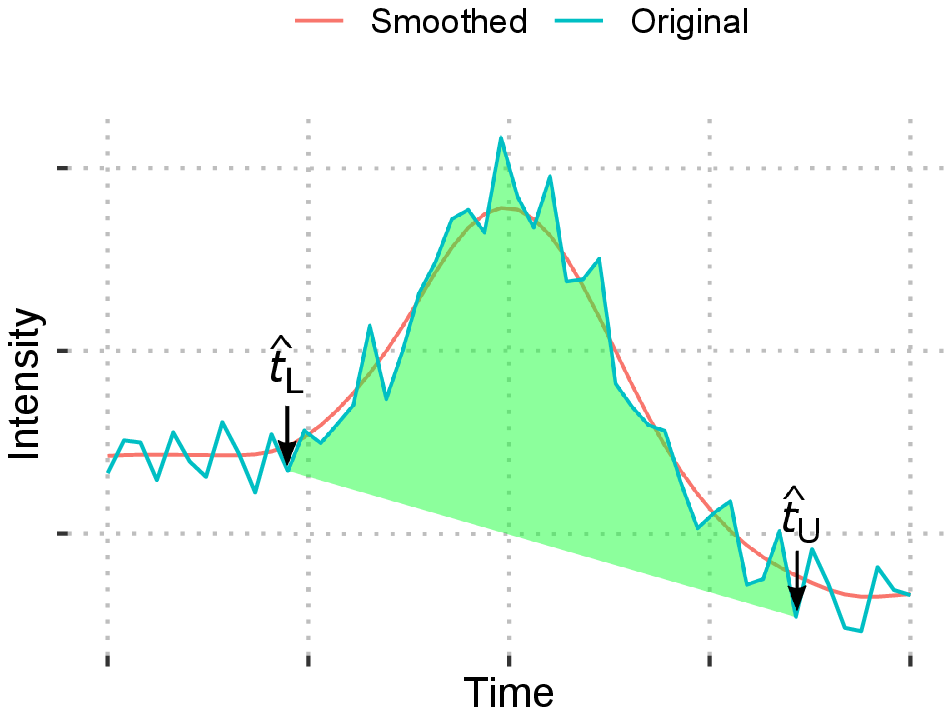}
        \caption{Peak area after partial convex hull boundary correction.}
        \label{fig:peakboundssteps3}
    \end{subfigure}
    \caption{This Figure illustrates changes in the area polygon defined by estimations for lower ($\widehat{t_\mathrm{L}}$) and upper ($\widehat{t_\mathrm{U}}$) peak boundaries after initial approximation (\ref{fig:peakboundssteps1}) and the two subsequent correction steps. First a Newtonian friction inspired approach is used in conjunction with the smoothed time series to widen the bounds (\ref{fig:peakboundssteps2}). The resulting potentially self-overlapping area polygon is then modified by first calculating the lower part of the convex hull and removing separate polygons after (\ref{fig:peakboundssteps3}).}
    \label{fig:peakboundssteps}
\end{figure}

Peak RT ($t_\mathrm{R}$) and boundaries ($t_\mathrm{L}$, $t_\mathrm{U}$) (see Figure~\ref{fig:chromacomp}) are estimated in multiple steps. Figure~\ref{fig:peakboundssteps} serves as a visual aid for picturing changes in the peak polygon in each step.

\paragraph{Peak retention time (RT)}

Non-maximum suppression (NMS) \citep{neubeck2006nms} is used to identify local maxima in the CWT coefficient matrix. NMS chooses points that have exclusively greater value neighbouring points. 
Local maxima in the CWT matrix correspond to low frequency components, peak components in this case, in the chromatogram data.
For this reason NMS returns a list of peak RT with associated time, scale and coefficient values, each representing a potential peak in the data.

\paragraph{Peak boundary estimation}

To determine the peak area, lower and upper bounds of the peak component are needed. These are approximated in two steps. 

While peak RT ($t_R$) is projected to a local maxima in the CWT coefficient matrix, approximate inflection points of the peak distribution should correspond to minima. The vector of CWT coefficients with the same scale as the previously determined maximum is inspected for the closest minima next to the maximum. This gives us estimates for lower and upper peak bounds.

\paragraph{Friction boundary correction}

As the wavelet transformation models uniformly distributed gaussian distributions and chromatographic peaks are often subject to skew, these approximate boundaries need to be corrected.

A Newtonian friction inspired approach is performed. Similar to a physical body gradually sliding down a hillside until a slope is reached at which friction stops it from moving, the initially estimated boundaries of the peak component are increased until a specified slope threshold is reached. This is detailed in Algorithm~\ref{alg:boundcorrect}. It is advised to normalize the threshold to the intensity range of each analyzed chromatogram.

\begin{algorithm}[ht]
    \caption{Peak boundary correction. The size function returns the size of a list.}
    \label{alg:boundcorrect}
    

\begin{algorithmic}[1]

\Require $Y$: Smoothed chromatogram data
\Require $p$: List of peak boundaries
\Require $t$: Threshold 

\State $t \gets (\Call{max}{Y} - \Call{min}{Y}) \cdot t$

\ForEach{$(u, l)\in p$}

\While{$u < \Call{size}{Y} - 1$} 
\If {$Y_u - Y_{u + 1} > t$}
\State $u\gets u + 1$
\EndIf
\EndWhile

\While{$l > 0$} 
\If {$Y_l - Y_{l - 1} > t$}
\State $l\gets l - 1$
\EndIf
\EndWhile

\EndFor

\end{algorithmic}

\end{algorithm}

\paragraph{Partial convex hull boundary correction}

At this point resulting peak polygons may overlap themselves at their lower and upper boundaries when there is significant noise in the data. To circumvent this and to mimic the behavior of a human rater the \emph{bottom} of the convex hull\footnote{When the points of the time series between temporal lower and upper boundaries are considered points on a time-intensity plane, the convex hull is the smallest convex polygon that contains them all.} of the peak data is calculated. 
A variation of Grahams scan~\citep[see][]{Graham1972} is implemented. It is described in Algorithm~\ref{alg:convex}.
Next, all areas of the polygon separated from the main peak area are removed.
Peak boundaries are then set to the minimum and maximum $x$ (time) value of the polygon vertices and peak RT to the maximum $y$ (intensity) value.

\begin{algorithm}[ht]
    \caption{Partial convex hull boundary correction. Modification of Grahams scan \citep[see][]{Graham1972}. The size function return the current size of the stack, push and pop are standard stack operations of adding and removing an element and orient returns the orientation $o$ (clockwise ($o > 0$), counterclockwise ($o < 0$) or colinear ($o = 0$)) of an ordered triplet. }
    \label{alg:convex}
\begin{algorithmic}[1]

\Require $W$: Empty stack
\Require $p$: Vector of points ordered by increasing $x$ coordinate

\State \Call{push}{$W, p_1$}
\State \Call{push}{$W, p_2$}

\For{$i \gets 3$ to $N$}

\While{$\Call{size}{W} \geq 2 \And \Call{orient}{p_i, W_{top}, W_{second}}\leq 0$} 

\State \Call{pop}{$W$}

\EndWhile

\State \Call{push}{$p$}

\EndFor

\end{algorithmic}

\end{algorithm}

\subsubsection{Peak area, height, background and slope}

Peak area ($a \in \mathbb{R^+}$), height ($h \in \mathbb{R^+}$), background ($b \in \mathbb{R^+}$) and slope ($s \in \mathbb{R}$) can be calculated after the peak boundaries are known (see Figure~\ref{fig:chromacomp}). Note that of these metrics only peak area and height are used in the evaluative section of this paper. Formulas for background and slope are described for completeness.

\paragraph{Area under the curve (AUC)} The AUC is defined as the area integral chromatographic data constraint to the peak component subtracted by the area integral of the slope between lower and upper peak bounds. This is equivalent to the peak polygon area and can easily be calculated using gauss's area formula:

\begin{equation}
\label{eq:auc}
  a=\frac{1}{2}{\Big |}\sum _{{i=1}}^{{n-1}}x_{i}y_{{i+1}}+x_{n}y_{1}-\sum _{{i=1}}^{{n-1}}x_{{i+1}}y_{i}-x_{1}y_{n}{\Big |} \,,
\end{equation}

where $x \in \mathbb{R^+}$ are the temporal and $y \in \mathbb{R^+}$ the intensity dimension coordinates of all measurements between peak bounds.

\paragraph{Peak slope} The slope $s \in \mathbb{R}$ is defined through the time positions of lower and upper peak boundaries:

\begin{equation}
    s = \frac{y(t_\mathrm{U})-y(t_\mathrm{L})}{t_\mathrm{U}-t_\mathrm{L}} \,,
\end{equation}

where $y(t) \in \mathbb{R^+}$ is the chromatogram time series and $t_\mathrm{U} \in \mathbb{R^+}$ is the time of the lower peak boundary.

\paragraph{Peak background} The background $b \in \mathbb{R^+}$ is determined by the temporal position of the peak maximum on the peak slope with the lower peak bound as the line intercept: 

\begin{equation}
    b = y(t_\mathrm{L})+(t_\mathrm{R}-t_\mathrm{L})\cdot s \,,
\end{equation}

where $t_\mathrm{L} \in \mathbb{R^+}$ is the time of the lower peak boundary.

\paragraph{Peak height} The height $h \in \mathbb{R^+}$ is defined as the difference of the peak maximum to its background:

\begin{equation}
    h = y(t_\mathrm{R})-b \,,
\end{equation}

where $t_\mathrm{R} \in \mathbb{R^+}$ the time of the peak maximum.

\subsubsection{Signal-to-noise ratio (SNR)}

While there is a lack of definitive guidelines for SNR measurements, signal is often defined as maximum peak height above baseline and noise as standard deviation or root mean squared of a manually selected part of the chromatogram where no peaks are present \citep{wells2011snr}. 

We propose to approximate SNR ($r \in \mathbb{R^+}$), by dividing standard deviation of the high-pass data ($H(x) \in \mathbb{R^+}$, mean $\overline{H}$, number of data points $N$) by doubled peak height ($h$):

\begin{equation}
  r = \frac{\sqrt{\frac{1}{N-1} \sum_{i=1}^N (H(x) - \overline{H})^2}}{2h}
\end{equation}

This method can be performed with very little or no data points outside of the peak bounds.
Note that this method of estimating SNR is dependent on the smoothing hyperparameter $\sigma$. This means that SNR can only be compared across analytes when the same $\sigma$ is used.

\subsubsection{Peak selection}
\label{sec:peakselection}

The peak identification procedure performed previously returns multiple possible peaks.

To select the appropriate one a peak fitness function

\begin{equation}
    f(t_{\mathrm{R}}) = c \cdot{} g(t_{\mathrm{R}})
\end{equation}

is calculated for each peak depending on its RT ($t_{\mathrm{R}} \in \mathbb{R^+}$), its CWT coefficient ($c \in \mathbb{R^+}$) and a RT fitness function ($g(t_{\mathrm{R}})$). The peak with the highest fitness value is selected and peaks with negative signed fitness are rejected completely. Note that this can lead to cases where no peak is detected in a chromatogram. These are considered non-detectables (ND).

The RT fitness function $g(t_{\mathrm{R}})$ is used for the selection of peaks depending on available information about the expected peak RT.
When no information about the expected RT is known, the RT fitness function is equivalent to $g(t_{\mathrm{R}}) = 1$.
This leads to selection of the peak with the highest CWT coefficient and rejects none. In other cases where an estimate of the RT ($\widehat{t_\mathrm{R}} \in \mathbb{R^+}$) and a range hyperparameter ($a \in \mathbb{R^+}$) is available, a quadratic fitness function

\begin{equation}
\label{eq:timefit}
g(t_{\mathrm{R}}) = 1 - \left(\frac{t_{\mathrm{R}} - \widehat{t_\mathrm{R}}}{a}\right)^2 \,,
\end{equation}

is created to weigh peaks by proximity to expected RT and leads to rejected peaks outside of the range.

\subsection{Whole dataset quantification}

A whole dataset is comprised out of multiple samples. Each sample contains data associated with multiple measured analytes. Besides the chromatogram by which an (unknown) analyte concentration is measured directly, it also contains a chromatogram of an internal standard analyte (IS) with a known analyte concentration.
The use of stable isotopically labelled internal standards is a commonly used technique that helps to control variability in quantitative assays \citep{dolan2012internalstandard,lcmsguide}.
There are also several calibration samples which are used for concentration calibration. These samples contain a known staggered amount of the analyte in addition to the known amount of IS.
The single chromatogram processing described in the previous section, is modified by hyperparameters specified for each analyte and IS. 

\begin{figure}[ht]
    \centering
    \includegraphics[width=\textwidth]{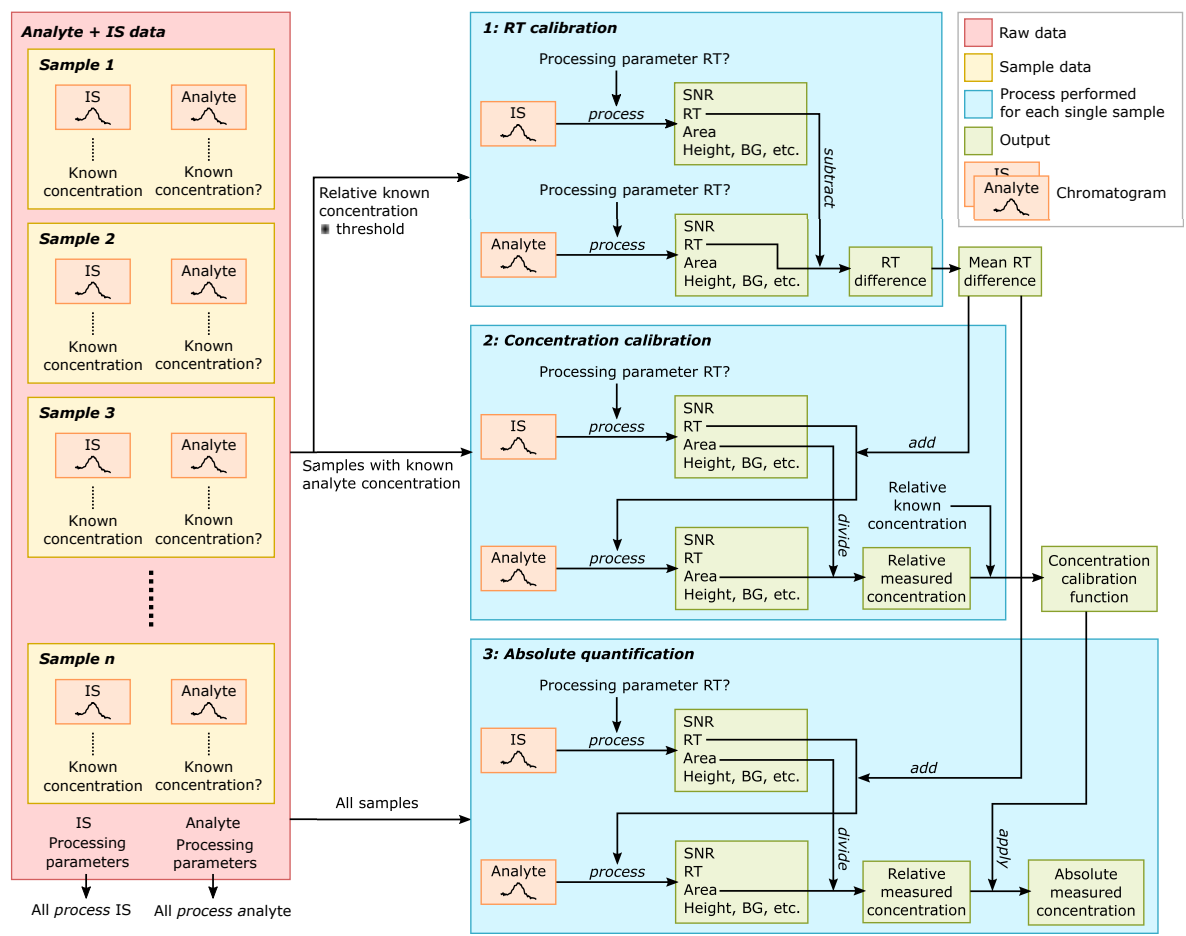}
    \caption{System diagram for a single analyte and its internal standard of a complete dataset. Samples contain an IS chromatogram with a known concentration and an analyte chromatogram which may have a known concentration. There are processing parameters for both analyte and IS which modify single chromatogram processing. RT calibration, concentration calibration and the final quantification are three separate steps which each process a different set of samples. Green boxes indicate outputs of processes.}
    \label{fig:system}
\end{figure}

The overall process of whole dataset quantification consists out of three main parts, RT calibration, concentration calibration, and final quantification. These are described in the following sections. Figure~\ref{fig:system} contains a system diagram which gives an overview over all parts of the routine.

\subsubsection{Retention time (RT) calibration}
\label{sec:rtcalibration}

While the RT of an analyte should be known approximately, it has some variance \citep{klammer2007improving,guo1987effects}.
The aim of this step is to analyze analyte chromatograms in which the right peaks are easily identifiable, to provide better estimates for the expected RT of a specific analyte and subsequently make peak selection easier in all other samples of the analyte.

Out of the calibration samples all are selected which contain a sufficient quantity of the calibrator analyte to be able to easily and distinctly detect the mass chromatogram peaks. Selected samples have a known concentration ratio equal or greater than a threshold hyperparameter $x$ ($C_\mathrm{A}/C_\mathrm{IS} \geq x$, where $C_\mathrm{A} \in \mathbb{R^+}$ is the known concentration of the analyte and $C_\mathrm{IS} \in \mathbb{R^+}$ is the known concentration of the IS).

Both IS and analyte chromatogram are processed as described in previous sections.
When wrong peaks are present in the data a guessed rough estimate of the expected RT and an acceptable range in the analyte parameters can be specified to create a preliminary RT fitness function (see~\ref{sec:peakselection}).

After obtaining all peak time positions of the selected samples, their arithmetic mean is used as an expected RT difference.

Now a calibrated RT for both analyte and internal standard has been calculated. The data showed that the standard deviation of both of them individually is consistently greater than the standard deviation of their difference (see results Table~\ref{tab:caldata}).
In subsequent analyses the IS peak RT ($t_\mathrm{R}^\mathrm{IS} \in \mathbb{R^+}$) is retrieved first. IS peaks are generally easier to identify because of their constantly high concentration. Then the previously calculated mean RT difference ($t_\Delta \in \mathbb{R}$) is used to calculate the expected RT of the analyte $t_\mathrm{R}^\mathrm{A} = t_\mathrm{R}^\mathrm{IS} + t_\Delta$.

In combination with a smaller acceptable RT range, this allows to build more accurate, dynamic RT fitness functions for the next chromatogram processing steps.

\begin{table}[ht]
    \centering
    
\begin{tabular}{lcccc} \toprule
Analyte & ${t^\mathrm{A}_\mathrm{R}}$ & ${t^\mathrm{IS}_\mathrm{R}}$ & ${t_\Delta}$ & $\beta$ \\ \midrule
Cortisol & 4.687 (0.018) & 4.682 (0.019) & 0.005 (0.007) & 2.41 \\ 
Cortisol Q & 4.688 (0.018) & 4.682 (0.019) & 0.006 (0.008) & 4.35 \\ 
Cortisone & 4.506 (0.014) & 4.493 (0.013) & 0.013 (0.007) & 1.61 \\ 
Testosterone & 5.841 (0.044) & 5.810 (0.040) & 0.036 (0.007) & 0.31 \\ 
Progesterone & 7.108 (0.055) & 7.042 (0.054) & 0.066 (0.006) & 1.07 \\ 
DHEA & 6.135 (0.046) & 6.098 (0.045) & 0.036 (0.007) & 0.76 \\ \bottomrule
\end{tabular}

    \caption{Calibration data for RT and concentration calibration. Measured RT mean and standard deviation (in parentheses) of the RT calibration samples (see section~\ref{sec:rtcalibration}). Note that the standard deviation of analyte RT (${t^\mathrm{A}_\mathrm{R}}$) and IS RT (${t^\mathrm{IS}_\mathrm{R}}$) is consistently greater then the standard deviation of their difference ${t_\Delta}$. $\beta$ is the slope of the concentration calibration regression model. Cortisol Q refers to the second most sensitive transition of cortisol, DHEA to dehydroepiandrosterone.}
    \label{tab:caldata}
\end{table}

\subsubsection{Concentration calibration}
\label{sec:conccalibration}

The calibration samples, as previously described, contain known concentrations of the analyte as well as IS.
While only calibration samples with a large known concentration ratio were used when calibrating the RT, now all of them are processed.

First, the IS chromatogram is processed like before. Then the analyte chromatogram is processed using the dynamic RT fitness function which depends on the IS RT and the mean RT difference calculated in the previous step.
Then the analyte peak area ($a_\mathrm{A} \in \mathbb{R+}$) is divided by IS peak area ($a_\mathrm{IS} \in \mathbb{R+}$) to obtain a relative measured concentration ($M=a_\mathrm{A}/a_\mathrm{IS}$). In combination with the relative known concentration ($C = C_\mathrm{A}/C_\mathrm{IS}$) available for each sample in this step, their linear relationship is modeled using linear regression without an intercept.

\begin{equation}
C = \beta M
\end{equation}

\begin{figure}[ht]
    \centering
    \includegraphics[width=0.35\textwidth]{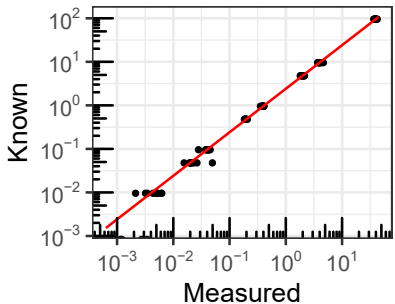}
    \caption{Plot of the linear regression model for concentration calibration of cortisol using our data.}
    \label{fig:curvecalexample}
\end{figure}

The resulting regression model can be used to estimate absolute concentrations from measured relative concentrations whenever the linearity assumption holds. See Figure~\ref{fig:curvecalexample} for a visual example of such a model.

\subsubsection{Absolute quantification}

Absolute sample quantification can now be performed for all samples. 
The chromatogram processing is performed just as in the previous, concentration calibration step.
To obtain absolute measured concentrations the concentration calibration regression function of the previous step is applied to each relative measured concentration.

\subsection{Software availability}
\label{sec:referenceimpl}

We supply\footnote{Code will be released upon acceptance of this paper.} a open source reference application which implements all parts of the algorithm. The software reads .mzML data, an open source standard for mass spectrometry data \citep{martens2010mzml}.

\section{Method}
\label{sec:exp}

\paragraph{Specimens Collection and Preprocessing}

The raw mass chromatogram data were made available from the Dresden longitudinal study of chronic stress and cognitive control (StressCog)~\citep[see][]{osfdata2019}. Study approval was granted by the Dresden ethics committee (dossier EK23012016, IRB00001473, and IORG0001076).
In the recruitment process 8,400 eligible participants from the population registry of the City of Dresden between 25 to 55 years old and within the metropolitan area, were selected and contacted by invitation letter. 
The study Protocol encompassed the collection of hair samples with a min lenght of 3 cm at the posterior vertex region of the head which was required for analyte extraction~\citep[see][]{gao2013quantitative}.
See~\citet[][]{schmidt2020reconsidering} for more details of the sampling and data collection process.

The analyses in this paper were performed on a subset of 230 hair samples from 229 different human subjects (137 females; age range 25 - 55; mean age = 38.05; age SD = 8.51) of the StressCog study. Samples were processed following the protocol described in \citet{gao2013quantitative} with slight adjustments. The non-pulverized samples were washed with isopropanol and steroid hormones were extracted from 7.5 mg hair by methanol incubation. Column switching on-line solid phase extraction (SPE) was performed, followed by analyte detection on an AB Sciex API 5000 QTrap mass spectrometer.
When enough material for two LC-MS/MS measurements was available the samples were split. This resulted in 446 samples which were processed by the LC-MS/MS (calibration samples excluded) to obtain selected-reaction monitoring (SRM) chromatograms. 

\paragraph{Analyte Quantitation}

The performance of the algorithm was evaluated by comparing its results to the manual quantification by two trained experts, referred to as expert 1 and expert 2. They are both trained in the manual marking of chromatographic peaks. Expert 1 was more experienced, and has manually marked approximately 60,000 samples with each 6 analytes before analyzing the data used in this paper, while expert 2 has marked about a tenth of this sample count, 5,000 samples, with 5 different analytes.

225 SRM chromatograms were analyzed by one expert, 221 by two experts. 216 of these LC-MS/MS samples were split from the same hair samples, which were therefore measured and analyzed twice. Multiplied by 6 different investigated analytes this results in 2676 individual measurements analyzed by one or both experts. All samples were also analyzed using the algorithm. These numbers are further visualized in supplementary Figure~\ref{fig:samplecounts}.

The measured analytes were cortisol, cortisone, testosterone, progesterone and dehydroepiandrosterone (DHEA). The selection of analytes and the procedure follows~\cite{gao2013quantitative}. The second most sensitive transition of cortisol (indicated by the Q suffix) was also included in the selection of analytes.

To be able to programmatically access the raw mass chromatograms, the data was converted from a proprietary file format to .mzML, using the ProteoWizard MSConvert software~\citep{kessner2008proteowizard,chambers2012cross}. Then the supplied software implementation (see~\ref{sec:referenceimpl}) was used to process the data.

For the manual quantification data, the AUC of both analyte and IS were marked and extracted by the experts using the AB SCIEX Analyst software version 1.6.2.

\paragraph{Algorithm settings}

Algorithm hyperparameters were mostly the same across analytes with the exception of fixed RT windows (expected IS RT and acceptable range) which had to be specified for the IS of progesterone, testosterone and DHEA, because of wrong IS peaks close to the correct one.

The absolute measured concentration (analyte AUC divided by IS AUC) was used for all evaluative statistics. Manual compiled relative concentrations were also scaled by the $\beta$ of the algorithms concentration calibration regression. This increases comparability between analytes for calculations which include all analytes in comparison to using the relative concentrations. And by not using the manually compiled values to create a separate calibration regression systematic errors are not overestimated.

\paragraph{Statistical analyses}

Statistical analyses were performed with R version 3.6 \citep{RCT2020}, latent variable models were estimated with lavaan \citep{Rosseel2012}.

\section{Results}

Table~\ref{tab:resanadesc} contains descriptive statistics for measured relative concentrations as well as SNR of each analyte.

\begin{table}[ht]
    \centering
    \small
    \begin{tabular}{lrrrrrrrrr} \toprule
Analyte & \multicolumn{1}{l}{N} & \multicolumn{4}{l}{Concentration (pg/mg)} & \multicolumn{4}{l}{SNR (Algorithm)} \\ 
~ & ~ & \multicolumn{2}{l}{Algorithm} & \multicolumn{2}{l}{Expert 1} & \multicolumn{2}{l}{Analyte} & \multicolumn{2}{l}{IS} \\ \midrule
Cortisol & 389 & 8.07 & (24.56) & 8.10 & (25.53) & 15.12 & (5.57) & 25.23 & (1.74) \\ 
Cortisol Q & 390 & 8.01 & (23.73) & 8.34 & (24.62) & 11.61 & (5.06) & 25.32 & (1.64) \\ 
Cortisone & 446 & 16.11 & (16.49) & 15.91 & (16.33) & 20.77 & (4.80) & 24.19 & (2.55) \\ 
Testosterone & 178 & 4.09 & (26.43) & 4.00 & (25.46) & 5.35 & (6.51) & 32.46 & (8.29) \\ 
Progesterone & 415 & 8.09 & (64.21) & 7.73 & (61.16) & 13.39 & (8.65) & 30.81 & (4.34) \\ 
DHEA & 414 & 1.89 & (2.60) & 1.85 & (2.59) & 13.74 & (6.52) & 34.52 & (3.34) \\ \bottomrule
\end{tabular}
    \caption{Descriptive statistics by analyte. Means followed by standard deviations in parentheses. All cases in which both the expert and the algorithm could detect a peak are included.}
    \label{tab:resanadesc}
\end{table}

\paragraph{Non-detectables (ND)}

Chromatograms where an expert cannot find a peak are marked as non-detectable (ND). In these cases, the analyte concentration is presumed to be less than the measuring instruments lower limit of detection. These cases were compared to cases when the algorithm cannot mark a peak with $\mathrm{AUC} > 0$ in the data.

Contingency data for successful peak marking is displayed in Table~\ref{tab:reslabels}. Sensitivity is close to $100\%$ across experts and analytes with the exception of testosterone where it is just over $95\%$. Specificity on the other hand is generally very low.
Testosterone had the highest rate of NDs with 261 of the 446 samples marked as ND by expert 1. The algorithms specificity for testosterone NDs was $54\%$.
Comparing all samples of all analytes measured by expert 1, there were 235 false positives.

\begin{table}[ht]
    \centering
    \small
    \begin{tabular}{lrrrrrrrrr} \toprule
    Analyte & \multicolumn{6}{l}{Expert 1 ($n=446$)} & \multicolumn{3}{l}{Accuracy (FP class.)}  \\
 ~ & \multicolumn{1}{l}{TP} & \multicolumn{1}{l}{FN} & \multicolumn{1}{l}{FP} & \multicolumn{1}{l}{TN} & Sensitivity & Specificity & A+S & AUC & SNR \\ \midrule
Cortisol & 389 & 2 & 16 & 39 & 99.49 \% & 70.91 \% & 1.00 & 0.98 & 0.87 \\ 
Cortisol Q & 390 & 0 & 45 & 11 & 100.00 \% & 19.64 \% & 0.91 & 0.92 & 0.75 \\ 
Cortisone & 446 & 0 & 0 & 0 & 100.00 \% & - & - & - & - \\ 
Testosterone & 178 & 7 & 121 & 140 & 96.22 \% & 53.64 \% & 0.95 & 0.88 & 0.95 \\ 
Progesterone & 415 & 0 & 30 & 1 & 100.00 \% & 3.23 \% & 0.88 & 0.88 & 0.81 \\ 
DHEA & 414 & 0 & 23 & 9 & 100.00 \% & 28.12 \% & 0.95 & 0.94 & 0.89 \\ \midrule
All & 2232 & 9 & 235 & 200 & 99.60 \% & 45.98 \% & 0.95 & 0.95 & 0.91 \\  \midrule 
    Analyte & \multicolumn{6}{l}{Expert 2 ($n=221$)} & \multicolumn{3}{l}{Accuracy (FP class.)}  \\
 ~ & \multicolumn{1}{l}{TP} & \multicolumn{1}{l}{FN} & \multicolumn{1}{l}{FP} & \multicolumn{1}{l}{TN} & Sensitivity & Specificity & A+S & AUC & SNR \\ \midrule
Cortisol & 199 & 4 & 4 & 14 & 98.03 \% & 77.78 \% & - & - & - \\ 
Cortisol Q & 205 & 2 & 10 & 4 & 99.03 \% & 28.57 \% & 1.00 & 1.00 & 0.93 \\ 
Cortisone & 220 & 0 & 1 & 0 & 100.00 \% & 0.00 \% & - & - & - \\ 
Testosterone & 106 & 5 & 45 & 65 & 95.50 \% & 59.09 \% & 0.98 & 0.86 & 0.97 \\ 
Progesterone & 217 & 0 & 4 & 0 & 100.00 \% & 0.00 \% & - & - & - \\ 
DHEA & 212 & 2 & 4 & 3 & 99.07 \% & 42.86 \% & - & - & - \\ \midrule
All & 1159 & 13 & 68 & 86 & 98.89 \% & 55.84 \% & 0.96 & 0.96 & 0.94 \\ \bottomrule
\end{tabular}
    \caption{
    Contingency data for peak detection. The algorithms sensitivity of peak detection was very high. Specificity was very low, mainly due to false positives (FP). Binomial regression was able to classify FPs (abbreviated FP class.) with high accuracy (area under the receiver operating characteristic). Equal group sizes were randomly sampled for hormones with at least 10 false positives. A+S indicates the regression model with both AUC and SNR as predictors.}
    \label{tab:reslabels}
\end{table}

\begin{figure}[ht]
    \centering
    \begin{subfigure}[t]{0.4\textwidth}
        \includegraphics[width=\textwidth]{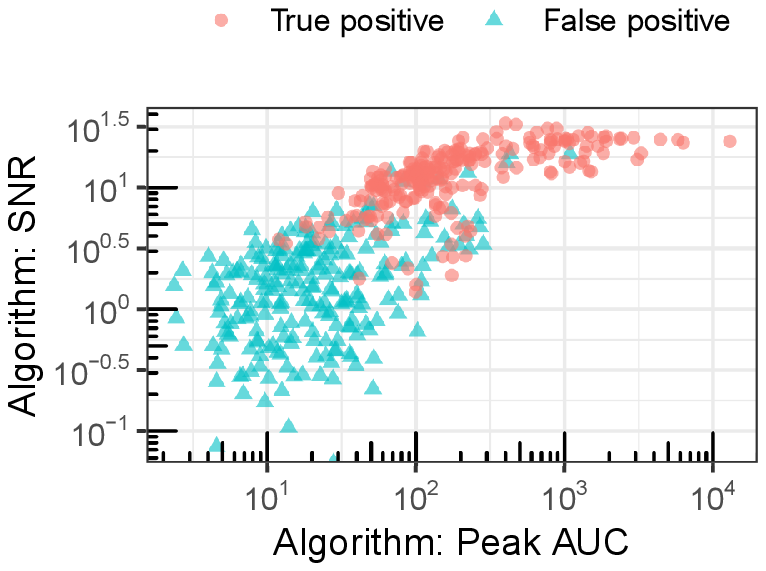}
        \caption{Scatterplot of expert 1 true and false positives.}
        \label{fig:ndlab}
    \end{subfigure}
    ~~
    \begin{subfigure}[t]{0.4\textwidth}
        \includegraphics[width=\textwidth]{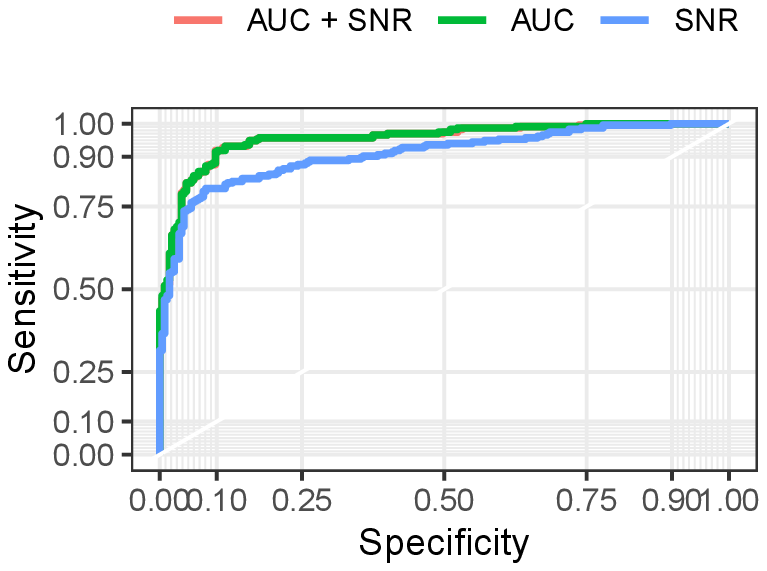}
        \caption{ROC for classifying expert 1 FPs. Accuracy (AUROC) of AUC: 0.95, SNR: 0.89, combined model: 0.95. The AUC model graph and the the combined model graph are oversecting.}
        \label{fig:roclab}
    \end{subfigure}
    \caption{Binominal regression classifiers for false positive peak detection. AUC and SNR as measured by the algorithm were investigated for their ability to predict false positives. These Figures show data of all analytes for expert 1. True positives were randomly sampled to obtain equal group sizes.}
    \label{fig:ndclass}
\end{figure}

It was investigated whether analyte peak AUC or analyte SNR could be used to predict false positives.
Three binomial regression models created. A combined model, $Y_\mathrm{A+S} = \beta_0 + \beta_1 X_1 + \beta_2 X_2$, where $Y$ is the probability of being an FP and $X_1$ and $X_2$ were AUC and SNR respectively, as well as models which just had AUC and SNR as predictors.
As the number of FPs was greatly lower than the number of TPs, the same number of TPs was randomly sampled to obtain equal group sizes. All three models were calculated for all analytes with at least 10 FPs as well as for the combined data. Overall accuracy (AUROC) for peak AUC was $0.91$, for SNR it was $0.95$ and the combined model had an accuracy of $0.95$. Accuracies for all calculated models are displayed in Table~\ref{tab:reslabels}.
Figure~\ref{fig:ndlab} is a scatterplot of the resampled cases with peak AUC and SNR on the axes.
Figure~\ref{fig:roclab} shows the receiver operating characteristic (ROC) for both dimensions.

\paragraph{Correlation \& differences}

Pearson correlation coefficients of the log transformed concentrations of all samples and analytes were all $r=0.98$ between algorithm and expert 1 and 2 as well as between the experts. Table~\ref{tab:rescor} shows single correlation coefficients for all analytes and raters. The supplementary Figure~\ref{fig:scatana} shows scatter plots for all analytes and the supplementary Figure~\ref{fig:ba} shows Bland-Altman plots of the differences between raters.

\begin{table}[!ht]
    \centering
    \begin{tabular}{lccc} \toprule
Analyte & Algorithm & Algorithm & Expert 1 \\ 
~ & Expert 1 & Expert 2 & Expert 2 \\ \midrule
Cortisol & 0.981 & 0.981 & 0.985 \\ 
Cortisol Q & 0.950 & 0.972 & 0.952 \\ 
Cortisone & 0.998 & 0.983 & 0.988 \\ 
Testosterone & 0.936 & 0.951 & 0.945 \\ 
Progesterone & 0.970 & 0.964 & 0.970 \\ 
DHEA & 0.953 & 0.898 & 0.952 \\ \midrule
All & 0.983 & 0.979 & 0.983 \\ \bottomrule
\end{tabular}
    \caption{Pearson correlation coefficients for log transformed hormone concentrations.}
    \label{tab:rescor}
\end{table}

\paragraph{Latent variable model}

The data contained several LC--MS/MS measurements of the same original samples, rated by the two experts and the algorithm (see supplementary Figure~\ref{fig:samplecounts}). An analogous structural equation model was created. Expert and Algorithm ratings were considered manifest variables, LC--MS/MS data was considered first order latent variables and \emph{true} hormone concentration as second order latent variable.

\begin{figure}[ht]
    \centering
    \includegraphics[width=0.35\textwidth]{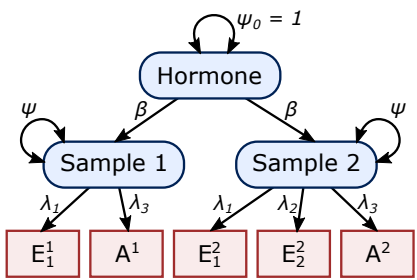}
    \caption{Path diagram of the measurement model. Rectangles represent manifest variables (Analyte concentrations estimated by experts $E_1$ and $E_2$ as well as by the algorithm $A$) for each LC--MS/MS sample. Rounded rectangles represent latent factors. First order latent factors are the measured samples, the second order latent factor is the hormone concentration. Residual variances are not shown in the diagram. }
    \label{fig:sem}
\end{figure}

Based on the latent state-trait theory \citep{steyer1999latent}, the total variance of analyte concentration per specimen replicate was decomposed into estimates of (1) reliability, (2) replicate specificity and (3) replicate consistency for each quantitation method (i.e., algorithm, rater 1, and rater 2). The model’s structure and its parameters are shown in Figure~\ref{fig:sem}. Additional method-specific variance components \citep[see][]{geiser2012comparison} were neither numerically identifiable nor necessary to account for the covariance among cortisol cortisone, and DHEA quantitations. The fit between the observed and the model-implied covariance structure was close for these analytes (\semres{10.417}{6}{0.108}{0.059}{0.019}). Accordingly, the algorithm measured the same construct as the human raters did. By contrast, the models for progesterone and testosterone quantitation showed a considerable lack of fit (\semres{21.632}{6}{0.001}{0.137}{0.307}) due to method-related residual covariance among the human raters, and replicate-related residual covariance among all methods, respectively. That is, the quantitation of progesterone by the algorithm differed slightly from the human raters, whereas there were differences in the within-replicate variance of testosterone irrespective of the quantatitation method. Reliability, specificity, and consistency of each quantitation method and analyte are reported in Table~\ref{tab:sem_est}.

\begin{table}[!ht]
    \centering
    \begin{tabular}{lrrrrrrrrr}
\toprule
  & \multicolumn{3}{l}{Consistency} & \multicolumn{3}{l}{Specificity} & \multicolumn{3}{l}{Reliability} \\
  & \multicolumn{1}{l}{$A$} & \multicolumn{1}{l}{$E_1$} & \multicolumn{1}{l}{$E_2$} & \multicolumn{1}{l}{$A$} & \multicolumn{1}{l}{$E_1$} & \multicolumn{1}{l}{$E_2$} & \multicolumn{1}{l}{$A$} & \multicolumn{1}{l}{$E_1$} & \multicolumn{1}{l}{$E_2$} \\
\midrule
Cortisol & 0.801 & 0.787 & 0.732 & 0.185 & 0.182 & 0.169 & 0.986 & 0.969 & 0.901\\
Cortisol Q & 0.760 & 0.762 & 0.708 & 0.206 & 0.207 & 0.192 & 0.966 & 0.968 & 0.900\\
Cortisone & 0.774 & 0.757 & 0.633 & 0.220 & 0.215 & 0.180 & 0.994 & 0.971 & 0.813\\
Testosterone & 0.869 & 0.896 & 0.699 & 0.102 & 0.105 & 0.082 & 0.971 & 1.002 & 0.782\\
Progesterone & 0.488 & 0.495 & 0.439 & 0.498 & 0.506 & 0.448 & 0.986 & 1.001 & 0.888\\
DHEA & 0.819 & 0.839 & 0.663 & 0.140 & 0.144 & 0.114 & 0.959 & 0.982 & 0.776\\
\bottomrule
\end{tabular}
    \caption{Consistency, specificity, and reliability of analyte quantitation (2nd replicate) by the algorithm ($A$), and both expert raters ($H_1$, $H_2$).}
    \label{tab:sem_est}
\end{table}

\section{Discussion \& Limitations}

Measurements automatically calculated by the algorithm correlate very strongly with those manually compiled by experts.
The algorithm returns far less non-detectables than expert raters. It was demonstrated that these peak detection false positives can be classified by the (algorithm measured) peak AUC and SNR. This leads us to recommend setting a minimum threshold for one or both of these values for each analyte with high ND rates (e.g. testosterone).

Latent variable models showed that the algorithm is able to measure the chromatograms trait very similar to the experts, while being closer to the more experienced experts results.
Consistency might be slightly overestimated due to list-wise exclusion.

Subsequent studies should experiment with modifications of algorithm parameters and the effect of disabling for example, the partial convex hull peak boundary correction which was designed to imitate expert rating behavior or use the smoothed peak area for calculating AUC. These modifications could potentially decrease model error.

Furthermore, it should be investigated if the algorithm is reliable with other types of data. These could be other LC--MS/MS mass chromatogram types as well as other types of chromatograms and spectrograms.
If nonlinear drift would emerge as a problem when analyzing different types of data, we suggest investigating the addition of a high-pass filter.

\subsection{Conclusion}

In conclusion, the algorithm presented here allows fast, automated, reliable and valid computational peak detection and quantification in LC--MS/MS and is also able to quantify SNR automatically.


\section{Acknowledgements}

Special thanks to the employees of Dresden LabService GmbH and Christian Rupprecht (Department of Engineering Science, University of Oxford) for their continuous support and suggestions. This work was supported by the German Research Foundation (DFG, Grant No. SFB 940/2).

\section{Conflict of interest}

The authors have no conflict of interest to declare.

\section*{References}

\bibliographystyle{elsarticle-num-names}
\bibliography{references.bib}

\newpage

\appendix
\section{Supplement}

\begin{table}[!hb]
    \centering
    
\begin{tabular}{@{}ll@{}}
\toprule
Abbreviation & Explanation \\ \midrule
AUC       & Area under the curve                           \\ 
AUROC     & Area under the ROC = Accuracy                  \\
IS        & Internal standard                              \\
LC--MS/MS & Liquid chromatography tandem mass spectrometry \\
ROC       & Receiver operating characteristic              \\
SNR       & Signal to noise ratio                          \\ \bottomrule
\end{tabular}

    \caption{Abbreviations}
    \label{tab:abbreviations}
\end{table}

\begin{figure}[ht]
    \centering
    \includegraphics[width=0.6\textwidth]{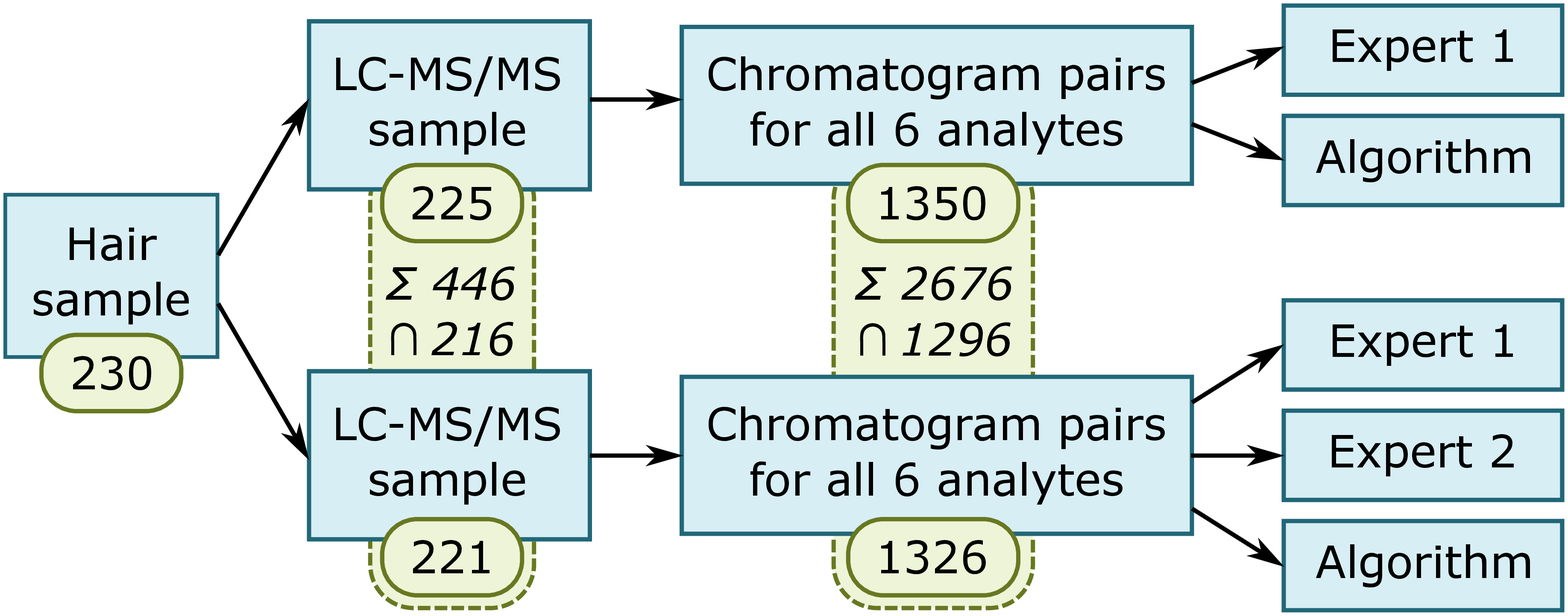}
    \caption{Visualization of sample number and dependence. 230 different hair samples were split and prepared to 446 LC-MS/MS samples. This Figure displays the resulting number of chromatograms and how many of them were repeated measurements. $\Sigma$ indicates the total sum including split samples and $\cap$ the number of intersecting original samples.}
    \label{fig:samplecounts}
\end{figure}

\begin{figure}[!hb]
    \centering
    \input{Fig10}
    \caption{Scatter plots of expert 1 and algorithm rating for each analyte concentration.}
    \label{fig:scatana}
\end{figure}

\begin{figure}[!hb]
    \centering
    \input{Fig11}
    \caption{Bland-Altman plots of log transformed measures. 95\% confidence intervals for the prediction of mean difference are indicated by the red lines. The blue line indicates mean difference. Note that the vertical axis is scaled using the hyperbolic arc-sine function to better depict differences of the skewed data.}\label{fig:ba}
\end{figure}


\end{document}